# Ferrielectric-paraelectric phase transitions in layered $CuInP_2S_6$ and $CuInP_2S_6$-$In_{4/3}P_2S_6$ heterostructures: A Raman spectroscopy and X-ray diffraction study

Rahul Rao,[1*] Ryan Selhorst,[1,2] Benjamin S. Conner,[3,4] Michael A. Susner[1]

[1]Materials and Manufacturing Directorate, Air Force Research Laboratory, 2179 12[th] Street, Wright-Patterson Air Force Base, Ohio 45433, USA

[2]UES Inc., Dayton, Ohio 45433, USA

[3]Sensors Directorate, Air Force Research Laboratory, 2241 Avionics Circle, Wright-Patterson Air Force Base, Ohio 45433, USA

[4]National Research Council, Washington, D.C. 20001, USA

## Abstract

$CuInP_2S_6$ (CIPS) is an emerging layered material that exhibits ferrielectric ordering well above room temperature (Curie temperature $T_C$ ~ 315 K). When synthesized with Cu deficiencies, CIPS spontaneously segregates into $CuInP_2S_6$ and $In_{4/3}P_2S_6$ domains (CIPS-IPS), which form self-assembled heterostructures within the individual lamellae. This re-structuring and resultant chemical pressure raises the Curie temperature and, depending on the Cu concentration, can be up to ~340 K for the highest Cu deficiency. In both CIPS and CIPS-IPS, the loss of polarization through the ferrielectric-paraelectric transition is driven by the movement of Cu ions within the lattice. Here we uncover the microscopic origins underpinning the phase transitions in pure CIPS and CIPS-IPS ($Cu_{0.4}In_{1.2}P_2S_6$) by performing a temperature-dependent Raman and XRD study. We measured the frequencies and linewidths of various cation and anion phonon modes and compared them to the extracted atomic positions from the refinement of XRD data. Our analysis shows that in addition to the Cu cation movement, the anion octahedral cages experience significant strains as they deform to accommodate the redistribution of Cu ions upon heating. This results in several discontinuities in peak frequencies and linewidths close to 315 K in CIPS. In the CIPS-IPS heterostructure, this process begins around 315 K and ends around 330 K.

---

[*] Correspondence – rahul.rao.2@us.af.mil

I. Introduction

Functional two-dimensional (2D) metal thio- and seleno-phosphates (collectively abbreviated here as MTPs) have recently come into focus due to their novel ferroelectric, ferromagnetic and multi-ferroic properties [1,2]. Some of these layered materials exhibit their properties down to the monolayer limits [3]; the prospect of combining functionalities of various layers in heterostructure devices, as well as uncovering new physics in moiré structures through rotational alignments makes MTPs very attractive candidates for next-generation functional devices [4–8]. The ferroic ordering exhibited in many MTP materials leads to temperature- and pressure-dependencies of these phase transitions; these are potentially tunable through dimensionality reduction, doping, strain etc. [1,9,10]. It is therefore important to study the microscopic origins of these phase transitions in order to engineer their properties for applications.

$CuInP_2S_6$ (CIPS), a room-temperature ferrielectric, has been well-studied of late due to its chemical stability and its high Curie temperature ($T_C$) of ~ 315 K in the bulk phase [11]. The CIPS crystal is monoclinic (space group $Cc$ in the ferrielectric state and C2/c in the paraelectric state) and each layer consists of $S_6$ octahedra circumscribing either metal cations or P-P pairs (Fig. 1a). The $S_6$ octahedra and the phosphorus atoms together form a structural backbone comprised of $[P_2S_6]^{4-}$ ethane-like anion groups that ionically pair with hexagonally arranged Cu and In cations. The metal cations may occupy three off-center sites within these octahedra as a result of a second-order Jahn-Teller instability associated with the $d^{10}$ electron configuration of Cu [12] - an off-center up or down site, a nearly-central site and a third, tetragonal site that extends into the interlayer van der Waals gap. Below $T_C$, the polar ordering of the Cu-sublattice (where they prefer the up ($Cu_1$) position within the octahedra) results in a small displacement of the $In^{3+}$ ions from the center of the lamellae an unequal distribution of the $Cu^+$ ions result in the predominant occupation of the up position within the octahedra while the $In^{3+}$ ions are displaced slightly downwards from the octahedral center. The up ($Cu_1$) and down ($Cu_2$) positions are denoted by the fully- and half-filled circles in Fig. 1a. This antiparallel displacement of the $Cu^+$ and $In^{3+}$ ions accounts for the room temperature out-of-plane ferrielectric polarization in CIPS [13].

Upon heating, the material becomes paraelectric above ~315 K. The transition has been found to be of the first order and is driven predominantly by Cu ion movement within the lamellae [14,15]. Above $T_C$, the $Cu^+$ ions equally occupy both the up and down positions within the $S_6$ octahedra. At higher temperatures, the occupation of the $Cu^+$ ions extends to the interlayer tetragonal sites, i.e. inside the van Der Waals gap; this results in a significant lattice expansion in addition to ionic conductivity [16].

When CIPS is synthesized with Cu deficiencies, the material spontaneously phase separates into ferrielectric $CuInP_2S_6$ and paraelectric $In_{4/3}P_2S_6$ (IPS) domains within the same single crystal with a common $[P_2S_6]^{4-}$ anionic foundation [17,18]. The IPS domains are embedded in a CIPS matrix and consist of an ordered arrangement of $In^{3+}$ ions and vacant sites in the octahedral network. Within the CIPS-IPS heterostructure, the IPS domains exert a chemical pressure over the CIPS domains, and the interplay between the two sub-lattices leads to an increase in the overall $T_C$ to ~335 K for highly Cu-deficient CIPS [17,18].

While the presence of a phase transition has been observed previously using various techniques such as calorimetry, X-ray diffraction (XRD), nuclear magnetic resonance (NMR) spectroscopy, Raman spectroscopy and dielectric measurements [14,15,19], there are open questions regarding the microscopic origins of the phase transitions in pure $CuInP_2S_6$ and in the $CuInP_2S_6/In_{4/3}P_2S_6$ heterostructures. For example, modeling suggests that the Jahn-Teller effect in CIPS is mediated by vibronic coupling, *i.e.* the coupling between *d* electrons in Cu and optical phonons [20]. Moreover, temperature-dependent optical absorption studies [21] have hinted at variations in the P-S and P-P bonds (and by extension their vibrational modes) due to the hopping of the $Cu^+$ ions within the lattice on heating up to $T_C$.

Electron-phonon coupling effects as well as lattice distortions can be measured using Raman spectroscopy. A previous temperature-dependent Raman study on CIPS single crystals showed preliminary evidence for coupling between $P_2S_6$ deformation modes and Cu ion vibrations [19]. However, a detailed analysis of temperature-dependent peak frequencies, intensities, and widths in CIPS and CIPS-IPS remains lacking. Temperature-dependent Raman scattering has also been used to study phase transitions in other layered MTP materials such as $CuCrP_2S_6$ [22], $CdPS_3$ [23] and $MnPS_3$ [24]. Here we present temperature-dependent Raman spectroscopy and XRD measurements on CIPS and CIPS-IPS (with the composition $Cu_{0.4}In_{1.2}P_2S_6$) single crystals. In particular, we focus on the anion vibrational modes (librations, deformation and stretching) and observe discontinuities in the frequencies, linewidths and intensities of several peaks in the vicinity of the phase transition temperatures. We correlate the trends in the vibrational modes to changes in the interatomic distances extracted from Rietveld refinement of the temperature-dependent XRD data. The loss of polarity above $T_C$ is accompanied by structural changes in both CIPS and CIPS-IPS, with the overall expansion of the unit cell. Local cation movements within the lamellae result in P-S, In-S, and S-S bond stretching, which are quantified by XRD data refinements and tied to changes in the Raman signatures.

## II. Methods

2.1 Synthesis and Chemical Characterization

The details of the crystal synthesis have been described elsewhere [17]. Briefly, single crystals of CIPS and CIPS-IPS were synthesized through vapor transport techniques. The precursor $In_2S_3$ (prepared from Alfa Aesar Puratronic elements, 99.999% purity, sealed in an evacuated fused silica ampoule and reacted at 950 °C for 48 hours) was reacted with the necessary quantities of Cu (Alfa Aesar Puratronic ), P (Alfa Aesar Puratronic ) and S (Alfa Aesar Puratronic ) to obtain the pure phase $CuInP_2S_6$ as well as a Cu-deficient composition $Cu_{0.4}In_{1.2}P_2S_6$. The starting materials were sealed in a fused silica ampoule with ~80 mg of $I_2$ and loaded into a tube furnace. The furnace was slowly ramped to 775°C over a period of 24 hours and held at that temperature for 100 hours. Afterwards, the samples were cooled at a rate of 20°C/hr. to ensure the growth of large domains.

## 2.2 Optical Characterization

Temperature-dependent Raman spectra were collected using a Renishaw inVia Raman microscope. The incident excitation (785 nm) was directed on to the sample through a Coherent Thz-micro low-frequency module (enabling measurements down to 10 cm$^{-1}$), and coupled to the Raman microscope with a fiber optic cable. A 50x magnification long-working distance objective lens was used to focus on to CIPS and CIPS-IPS crystals mounted within a temperature stage (Microptik). The laser power was set to a few mW to avoid additional heating from the laser. Raman spectra were collected with 15 s acquisition times and 4 accumulations. Spectral analysis was performed (in Igor Pro) by cubic spline baseline subtraction and Lorentzian peak fitting to extract frequencies and linewidths.

## 2.3 Structural Characterization

High-resolution synchrotron data from CIPS were obtained at beamline 11-ID-C at the Advanced Photon Source at Argonne National Laboratory. The samples were ground into a fine powder and carefully placed in a Cu cylinder to avoid issues related to preferential orientation in this layered material. This cylinder was in turn placed in a Linkam THM600 microscope stage and purged with inert gas. X-ray patterns were taken from low temperatures to high (with $\Delta T < 0.2$ K). The patterns were collected in transmission mode using a Perkin-Elmer large area detector; the wavelength of the synchrotron radiation was 0.117418 Å. The collected 2D patterns were processed into conventional 1D patterns (Fit2d software) which were then refined using the FullProf suite [25].

## III. Results and Discussion

Representative unpolarized, room temperature Raman spectra (collected with 785 nm laser excitation) from CIPS and CIPS-IPS are shown in Fig. 1b. The fitted peaks are shown below the raw data and the overall fitted spectra are overlaid on top. Both spectra exhibit several peaks between 10 – 650 cm$^{-1}$ that consist of external (<150 cm$^{-1}$) and internal (>150 cm$^{-1}$) vibrations. These vibrational modes can be divided into five frequency ranges, and in general are common to all layered MTPs. In the case of CIPS, the lowest frequency peaks (<50 cm$^{-1}$) and those between 50 and 150 cm$^{-1}$ correspond to cation and anion librations, respectively. Among the cation vibrational modes, the lower and higher frequency peaks correspond to vibrations from Cu and In, respectively. The peaks between 150 – 200 cm$^{-1}$ and 200 – 350 cm$^{-1}$ correspond to deformations of the S-P-P and S-P-S bonds [$\delta$(SPP) and $\delta$(SPS)] within the octahedra, respectively (or collectively, anion deformation modes). Some of the anion deformation modes are sensitive to the type of metal cation present and its location within its S$_6$ octahedron. In particular, the peak around 320 cm$^{-1}$ can be attributed to distortions within the S$_6$ cage occupied by the Cu$^+$ ions [19]. The high intensity peak around 370 cm$^{-1}$ is attributed to P-P stretching [$\nu$(PP)], and the PS$_3$ stretching modes [$\nu$(PS$_3$)] appear around 450 and 550 cm$^{-1}$. In the range of modes between 500 – 600 cm$^{-1}$, the lower (higher) frequency modes are influenced by Cu (In) cations [26,27].

The spectrum from CIPS-IPS also exhibits peaks grouped in the same frequency ranges described above, but is much more complex than the CIPS spectrum and we resolve 27 peaks as seen in Fig. 1b. The larger number of peaks in the CIPS-IPS Raman spectrum can be attributed to the CIPS and IPS sub-lattices within the material. A comparison of the CIPS-IPS Raman spectrum with spectra from the pure phase CIPS and IPS (Ref. [28]) enables us to identify peaks unique to both sub-lattices. In the low-frequency region (<90 cm$^{-1}$), we observe two groups of peaks. Similar to CIPS, the lower and higher frequency peaks within this frequency range can be attributed to Cu and In vibrational modes, respectively. Above 100 cm$^{-1}$, there are four peaks in the anion libration region between 100 and 140 cm$^{-1}$, two of which can be attributed to CIPS (100 and 114 cm$^{-1}$) and the other two to IPS (127 and 140 cm$^{-1}$). Similarly, between 200 and 300 cm$^{-1}$, the two most intense peaks appear at ~255 and ~270 cm$^{-1}$, and can be assigned to IPS and CIPS, respectively. A peak at 300 cm$^{-1}$ in the CIPS-IPS spectrum does not appear in the CIPS spectrum, thus it can be attributed to anion deformation in IPS. Furthermore, as previously mentioned, the peak at 320 cm$^{-1}$ can be assigned to anion deformation in CIPS. Among the anion stretching modes, the two highest frequency peaks around 580 and 610 cm$^{-1}$ do not appear in the CIPS spectrum and therefore can be attributed to IPS.

Next, we describe the temperature dependence of the Raman peaks in CIPS and CIPS-IPS, focusing on the vicinity of the two Curie temperatures ($T_C$ ~ 315 and 330 K for Cu$_0$InP$_2$S$_6$ and Cu$_{0.4}$In$_{1.2}$P$_2$S$_6$ [17], respectively). As described above, the loss of polarity across the ferrielectric-paraelectric phase transition is primarily driven by the movement of the Cu$^+$ ions as they redistribute between the up and down positions within the S$_6$ octahedra. This can be observed in the changes in the low-frequency cation vibrational modes. Fig. 2a shows a waterfall plot with Raman spectra between 300 and 350 K from a CIPS crystal. Here we focus on the low-frequency range (10 – 90 cm$^{-1}$), which exhibits two peaks, with the lower (higher) frequency peak corresponding to the Cu (In) extended translational modes. Temperature-dependent Raman spectra over the full measured range (10 – 650 cm$^{-1}$) are included in the Supplemental Material, Figs. S1 and S2 [29]. The low frequency peak centered around 35 cm$^{-1}$ in the room temperature spectrum can be deconvoluted into three Lorentzian peaks; the individual peaks are shown below the raw data, with the overall fit overlaid on the raw spectra. The three low-frequency modes appear around 25, 33 and 39 cm$^{-1}$ at room temperature, and are attributed to Cu$^+$ ion translations within the lattice. The peak around 70 cm$^{-1}$ is likewise attributed to translations of In$^{3+}$ ions [19].

Typically, all Raman vibrational modes exhibit redshifts in peak frequencies as well as broadening with increasing temperature due to lattice anharmonicity and thermal expansion [30,31]. In the case of the three Cu$^+$ modes in CIPS, we see either an anomalous blueshift in frequencies or they remain constant upon heating until the $T_C$ is reached, around 315 K. Above $T_C$, these peak frequencies redshift slightly with increasing temperature. This anomalous trend in Cu peak frequencies in CIPS has been observed previously and is discussed in Ref. [19]. Interestingly, however, our data reveals a change in peak intensities across $T_C$ that has not been previously reported. With increasing temperature, we see an overall redshift of the spectral lineshape with the lowest frequency peak (~25 cm$^{-1}$ at 300 K) increasing in intensity. In addition, the peak around 39 cm$^{-1}$ drops sharply in intensity and disappears above $T_C$. These trends are shown in Fig. 2b, which plots the ratio of intensity of the peak around 33 cm$^{-1}$ to the peak at 25 cm$^{-1}$ ($I_{33}/I_{25}$, left axis in Fig. 2b) against temperature. Also plotted in Fig. 2b are vertical dashed lines denoting the two Curie temperatures at 315 and 330 K for CIPS and CIPS-IPS, respectively. The $I_{33}/I_{25}$ ratio steadily decreases with temperature upon heating, and, interestingly, levels off above 330 K. The intensity of the 39 cm$^{-1}$ peak, on the other hand, drops sharply with increasing temperature and the peak cannot be resolved above 320 K. These dramatic changes in peak intensities around $T_C$ reflect the significant changes within the CIPS lattice due to Cu$^+$ ion movement/redistribution, eventually resulting in a loss of ferrielectric polarization above 315 K. Since the In$^{3+}$ ions do not drastically shift across the lamella like the Cu$^+$ ions across the phase transition temperature, we do not expect to see any anomalous jumps in frequencies or intensities in the In vibrational mode at 70 cm$^{-1}$. As expected, it exhibits a monotonic redshift in frequency and broadening with increasing temperature, without any anomalous change in peak intensity. Note that the In$^{3+}$ ions do move in order to accommodate the higher occupation of the Cu$^+$ ions in the down positions across $T_C$. This affects the In-S bond distances, and will be discussed later.

The corresponding temperature-dependent low-frequency Raman spectra from CIPS-IPS are shown in Fig. 2c. As described above, these spectra exhibit many more peaks compared to CIPS-IPS due to the two sub-lattices within the lamellae. Other than the expected temperature-dependent redshifts in peak frequencies, the lowest frequency peaks, which correspond to the Cu vibrational modes, do not exhibit any anomalous trends across the $T_C$ in CIPS-IPS (330 K). We also do not see any significant variations in peak intensities like those observed in CIPS (Fig. 2b). However, one peak stands out – the intensity of a vibrational mode at 70 cm$^{-1}$ decreases sharply around 330 K relative to neighboring peaks. The trend is shown in Fig. 2d, which plots the ratio of the intensity of the 70 cm$^{-1}$ peak to the neighboring peak at 76 cm$^{-1}$ ($I_{70}/I_{76}$). The $I_{70}/I_{76}$ ratio experiences a sharp dip around 330 K, and increases slightly above $T_C$. The origin of this peak is unknown; however, by comparing with the CIPS Raman spectrum (Fig. 1b), we can tentatively attribute the 70 cm$^{-1}$ peak to extended vibrations due to the In$^{3+}$ ions in the IPS lattice. In our previous study we showed that the lattice parameters of the CIPS sub-lattice increase sharply across the phase transition temperatures, and the most significant increase occurs for the *c* lattice parameters, *i.e.* perpendicular to the layers [18]. A discontinuity in the *c* lattice parameter of the IPS sub-lattice can also be observed around 330 K, reflecting interfacial effects between the two chemical phases. These effects are likely responsible for our observed intensity variations for the 70 cm$^{-1}$ peak.

While the loss of polarization can be primarily attributed to Cu$^+$ ion occupancy, their redistribution on heating also affects the surrounding P and S bonds within the octahedra. These effects should be observable through the anion deformation and stretching modes. To this end, we analyzed the temperature-dependent frequencies and widths (full-width at half-maximum intensity) of several anion vibrational modes in CIPS and CIPS-IPS. Below, we discuss the temperature-dependences of three peaks, which are the most prominent features in the Raman spectra. As labeled in Fig. 1b, they correspond to S-P-S deformations [$\delta$(SPS)], around 260 cm$^{-1}$ in CIPS and CIPS-IPS), P-P stretching [$\nu$(PP), around 375 cm$^{-1}$] and the high-frequency PS$_3$ stretching mode [$\nu$(PS$_3$), around 560 cm$^{-1}$]. Fig. 3 plots the temperature dependence of the frequencies and widths of these three peaks. In all the figures, we show the two Curie temperatures for CIPS (315 K) and CIPS-IPS (330 K) as dashed vertical lines. Note that the Lorentzian peak fitting revealed the peaks in CIPS-IPS to consist of two sub-peaks that can be attributed to contributions from the CIPS and the IPS sub-lattices. Owing to the closeness in frequencies, and being consistent with the assignments of the low-frequency cation vibrations, we assign the lower and higher frequency peak to the CIPS and IPS sub-lattices in CIPS-IPS, respectively.

The temperature-dependence (between 300 and 350 K) of the frequencies and widths of the S-P-S deformation [$\delta$(SPS)] modes are shown in Figs. 3a and 3b, respectively. As expected, all peaks exhibit an anharmonic redshift in frequency with increasing temperature. In addition, the $\delta$(SPS) mode in CIPS (red triangles in Fig. 3a) exhibits a clear and sharp decrease in frequency at 315 K, and continues to redshift up to 350 K. On the other hand, we see a discontinuity (albeit subtle) in the frequency of the CIPS peak from CIPS-IPS (black circles in Fig. 2a) at 330 K. The sharp decrease in the CIPS peak frequencies across $T_C$ can be attributed to tensile strains developed within the lamellae as the S$_6$ octahedra deform while trying to accommodate the movement and equal occupancy of the Cu$^+$ ions in the up and down positions within the octahedra. Concomitant to the frequency of the δ(SPS) mode in CIPS, the width also exhibits a sharp narrowing above 315 K (Fig. 3b). And unlike the subtle discontinuities in the frequencies of the δ(SPS) modes of CIPS-IPS, the peak widths of the two δ(SPS) modes in CIPS-IPS exhibit noticeable discontinuities at 330 K (Fig. 3b), with the lower frequency CIPS sub-lattice peak experiencing peak sharpening and the higher frequency IPS sub-lattice peak experiencing peak broadening. Typically, peak sharpening and broadening are associated with an increase or loss in crystallinity, respectively. The contrasting effects observed in CIPS-IPS could therefore be related to the deformations experienced by the sub-lattices as they exert chemical pressure on each other upon heating.

The effect of $Cu^+$ ion redistribution on the chalcogen backbone is perhaps clearest in the temperature-dependent frequency and width of the PP stretching mode [ν(PP)] as shown in Figs. 3c and 3d. The ν(PP) mode in CIPS (red triangles in Fig. 3c) exhibits a sharp increase in frequency just above 315 K, followed by anharmonic redshift with increasing temperature up to 350 K. The sharp increase can be attributed to a shortening of the P-P bond owing to compressive stresses as the $Cu^+$ ions redistribute between the up and down positions within the octahedra. In CIPS-IPS, we observe a blueshift of the ν(PP) mode at 315 K, which continues up to the $T_C$ of CIPS-IPS (330 K), followed by redshifted frequencies. This is seen for both of the ν(PP) modes in the CIPS and IPS sub-lattices (black and blue data in Fig. 3c, respectively). The steady increases in ν(PP) peak frequencies between 315 and 330 K suggest compressive strain (or shortening) in the P-P bonds in both CIPS and IPS sub-lattices. On heating, this strain begins at the $T_C$ of CIPS and ends at the $T_C$ of CIPS-IPS. This shows that, even though the loss of polarization occurs around 330 K in CIPS-IPS, structural changes begin at a lower temperature. Since CIPS-IPS consists of CIPS and IPS sub-lattices, the P-P bonds within the octahedra apparently shorten continuously as the sub-lattices expand, culminating in the transition around 330 K. The widths of all three peaks (Fig. 3d) exhibit broadening with temperature, with a slight discontinuity observed in the width of the CIPS peak across its $T_C$ at 315 K (red open triangles in Fig. 3d). The width of the ν(PP) peak from the IPS sub-lattice in CIPS-IPS exhibits an anomalous decrease between 315 and 330 K (blue open circles in Fig. 3d). This behavior could be attributed to changes in crystallinity due to the restructuring of the CIPS and IPS sub-lattices with increasing temperature.

In the high frequency region, the $\nu(PS_3)$ peak in CIPS increases in frequency up to $T_C$, above which it redshifts with temperature (Fig. 3e). The width of the $\nu(PS_3)$ peak experiences a significant jump across $T_C$ (Fig. 3f), followed by a steady increase with temperature. These trends suggest significant strains in the P-S bonds, likely as a result of increasing $Cu^+$ ion movement with temperature. In the case of CIPS-IPS, the effects of temperature on the $\nu(PS_3)$ peaks are subtler. The $\nu(PS_3)$ peak frequencies in both CIPS and IPS sub-lattices exhibit slight discontinuities around 330 K (Fig. 3f), while the width of the $\nu(PS_3)$ peak in IPS exhibits an anomalous decrease between 315 and 330 K (Fig. 3f), similar to the width of the $\nu(PP)$ peak from the IPS sub-lattice (Fig. 3d). In addition to trends in the frequencies and widths, we also observed changes in intensities of some of the anion vibrational modes. In CIPS the anion deformation mode at 320 $cm^{-1}$ and libration mode at 100 $cm^{-1}$ experience a sharp decrease in intensity with increasing temperature. Above $T_C$, these intensities become constant. A similar behavior is seen in CIPS-IPS for the high frequency mode at 558 $cm^{-1}$ [$\nu(PS_3)$ peak in the CIPS sub-lattice], which decreases in intensity relative to the mode at 565 $cm^{-1}$, as well as the anion libration mode at 100 $cm^{-1}$. These intensity trends are included in the Supplemental Material, Fig. S3.

Further insights into the anomalous frequency and width trends presented in Fig. 3 can be obtained from temperature-dependent XRD data. Crystallographic data were obtained by performing Rietveld refinements on temperature-dependent XRD patterns collected upon heating CIPS powder samples. Some of these XRD data have been published previously in Ref. [18]. From these data, we extracted the distances between the various S atoms in the octahedra, as well as P-S and In-S distances. To make the refinements stable, we were forced to fix the P positions in this analysis. Nonetheless, the other data provide valuable insights into the trends observed in the Raman peak frequencies and widths. The temperature-dependent (between 295 – 350 K) S-S, P-S and In-S distances within a single octahedron in a lamella are presented in Fig. 4, with the data points corresponding to the top and bottom of the octahedra plotted as open and filled data, respectively. The corresponding atom numbering ($S_1$, $P_1$ etc.) is shown in Fig. 1a.

Fig. 4a plots the distances between the S atoms at the top and bottom of the octahedra. From the figure, we see that the S-S distances at the top of the octahedra, i.e. $S_1$-$S_2$, $S_2$-$S_3$ and $S_1$-$S_3$ exhibit increases up to $T_C$, after which they remain constant up to 350 K. At the same time, the S-S distances at the bottom of the octahedra do not experience significant changes. These differences in the S-S distances at the tops and bottom of the octahedra can be attributed to the movement of the $Cu^+$ ions from fully occupied "up" positions below 315 K to an equal distribution between the "up" and "down" sites above this temperature, culminating in the loss of polarization above $T_C$. As the $Cu^+$ down occupancy increases, we also see dramatic changes in the P-S distances. The data in Fig. 4b show that the $P_2$-S distances at the bottom of the octahedra increase significantly above $T_C$ and up to ~340 K. On the other hand, the $P_1$-S distances at the tops of the octahedra decrease above $T_C$. The trends in the P-S interatomic distances indicate an elongation of the lamellae towards the bottom of the octahedra in order to accommodate the increasing occupancy of the $Cu^+$ ions with temperature. In the ferrielectric phase, the $In^{3+}$ ions are displaced slightly downwards from the octahedral center (Fig. 1a). With an increase temperature, the In-S distances at the bottom of the octahedra increase across $T_C$, while those at the top appear to decrease (Fig. 4c). The increase in the In-S distances at the bottom of the octahedra are a result of an upward displacement of the $In^{3+}$ ions closer to the octahedral centers. This occurs in order to accommodate the $Cu^+$ ions in the down positions.

The data presented in Fig. 4 show that significant restructuring and strains are imposed on the octahedra to accommodate the increased movement of the $Cu^+$ ions as they fill the down positions within the $S_6$ octahedra and eventually extend outwards into the van der Waals gap at higher temperatures. The restructuring includes shortening and elongation of the P-S bonds at the top and bottom of the octahedra, as well as variations between the S-S distances due to distortion of the octahedra. These distortions impose tensile strain on the S-P-S bonds, resulting in the observed sharp redshift in the frequency of the $\delta$(SPS) peak (Fig. 3a). The concomitant decrease in $\delta$(SPS) peak width (Fig. 3b) could be attributed to a temporary increase in lattice crystallinity brought about by the restructuring of the $S_6$ octahedra and the occupancy of the $Cu^+$ ions in the down positions. At the same time, the elongation of the octahedra result in tensile and compressive strains in the P-S bonds, culminating in the observed blueshift in the ($PS_3$) stretching mode (Fig. 3e) and large increase in its width (Fig. 3f) across the ferrielectric-paraelectric transition temperature. The variations in the P-S distances are also likely responsible for the compression of the P-P bonds and the striking trends observed in the frequencies and widths of the ν(PP) peak (Figs. 3c and 3d). Here, we have only presented temperature-dependent XRD data from CIPS. Owing to the much more complex crystal structure of CIPS-IPS, *i.e.* CIPS and IPS sub-lattices, and the shrinking and expansion of their domains upon heating [18], extraction of the temperature-dependent interatomic distances in CIPS-IPS is a significantly challenging task. Nonetheless, the similarities in the Raman spectral trends between CIPS and CIPS-IPS suggest that our analysis would be valid for CIPS-IPS as well.

**IV Conclusion**

In this work, we studied the ferrielectric-paraelectric phase transitions in CIPS and CIPS-IPS using Raman spectroscopy. Analysis of the temperature-dependent spectra revealed several structural changes that accompany Cu+ ion redistribution and that result in a loss of polarization above $T_C$ (~315 K). In CIPS we observed the intensity modulations in the low-frequency Cu vibrational modes, as well as anomalous jumps in the frequencies and linewidths of the anion vibrations. These trends were correlated with the interatomic distances obtained from temperature-dependent XRD data, and showed elongation of the S6 octahedral cages and significant strains in the S-S and P-S bonds. In CIPS-IPS observed changes in the Raman spectra beginning around 315 K and extending to its $T_C$ around 330 K. These results show that, while loss of polarization occurs above 330 K in $Cu_{0.4}In_{1.2}P_2S_6$, structural changes begin at a lower temperature. Our combined Raman spectroscopy and XRD study revealed a microscopic view into the ferrielectric-paraelectric phase transition processes in CIPS and CIPS-IPS. Furthermore, our work also hints at approaches to manipulate lattice strains and thereby ferroelectric behavior in CIPS and other related functional 2D materials.

**Acknowledgements**

We acknowledge support through the United States Air Force Office of Scientific Research (AFOSR) LRIRs 19RXCOR052 and 18RQCOR100, AOARD-MOST Grant Number F4GGA21207H002 and the National Research Council Postdoctoral Fellowship award. Use of the Advanced Photon Source, an Office of Science User Facility operated for the U.S. Department of Energy (DOE) Office of Science by Argonne National Laboratory, was supported by the U.S. DOE under Contract No. DE-AC02-06CH11357.

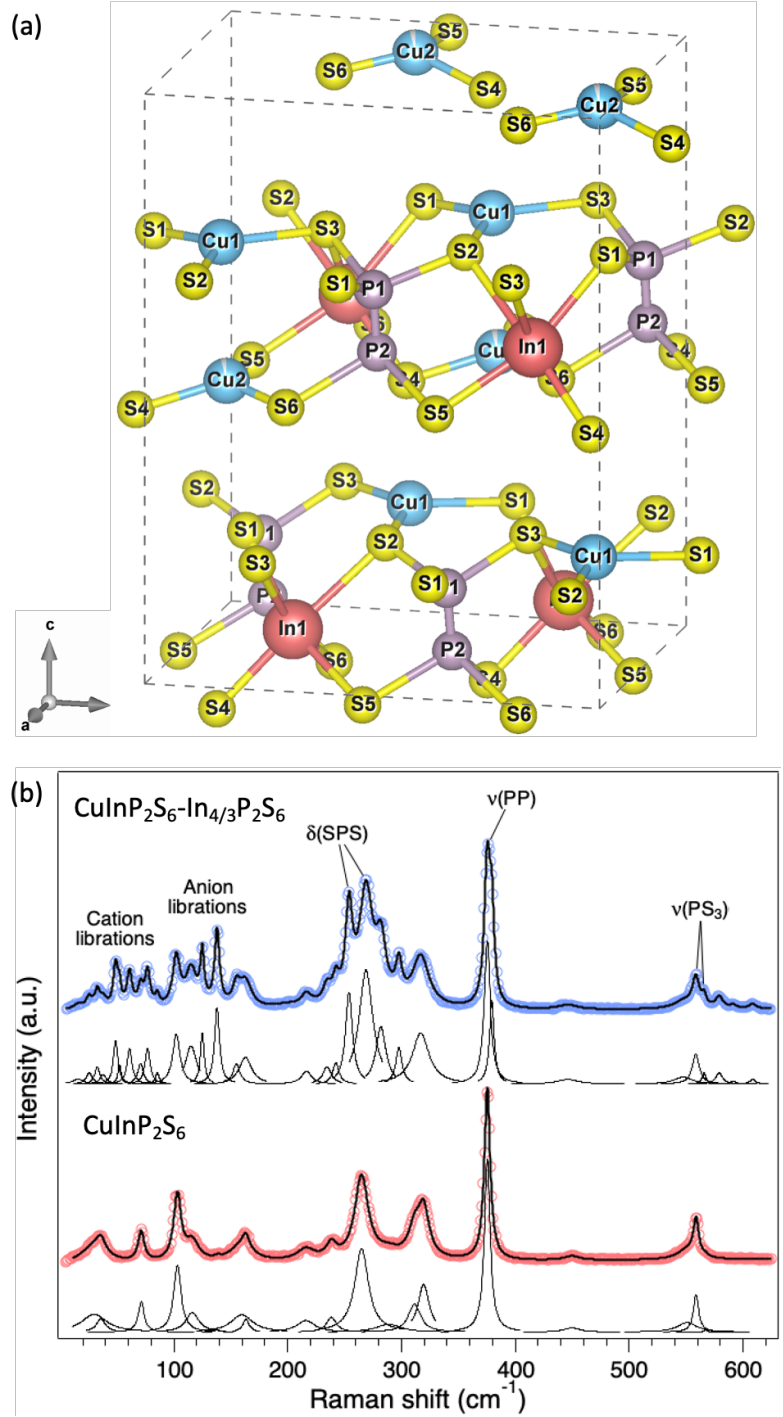

FIG. 1. (a) Schematic of the unit cell of CuInP$_2$S$_6$, with all the atoms labeled. (b) Room-temperature Raman spectra (785 nm excitation) from CuInP$_2$S$_6$ (CIPS, bottom spectrum) and CuInP$_2$S$_6$-In$_{4/3}$P$_2$S$_6$ (CIPS-IPS, top spectrum). The spectra were fitted to Lorentzian peaks and are plotted under the raw data (open circles) along with the overall fit (solid line) overlaid on top.

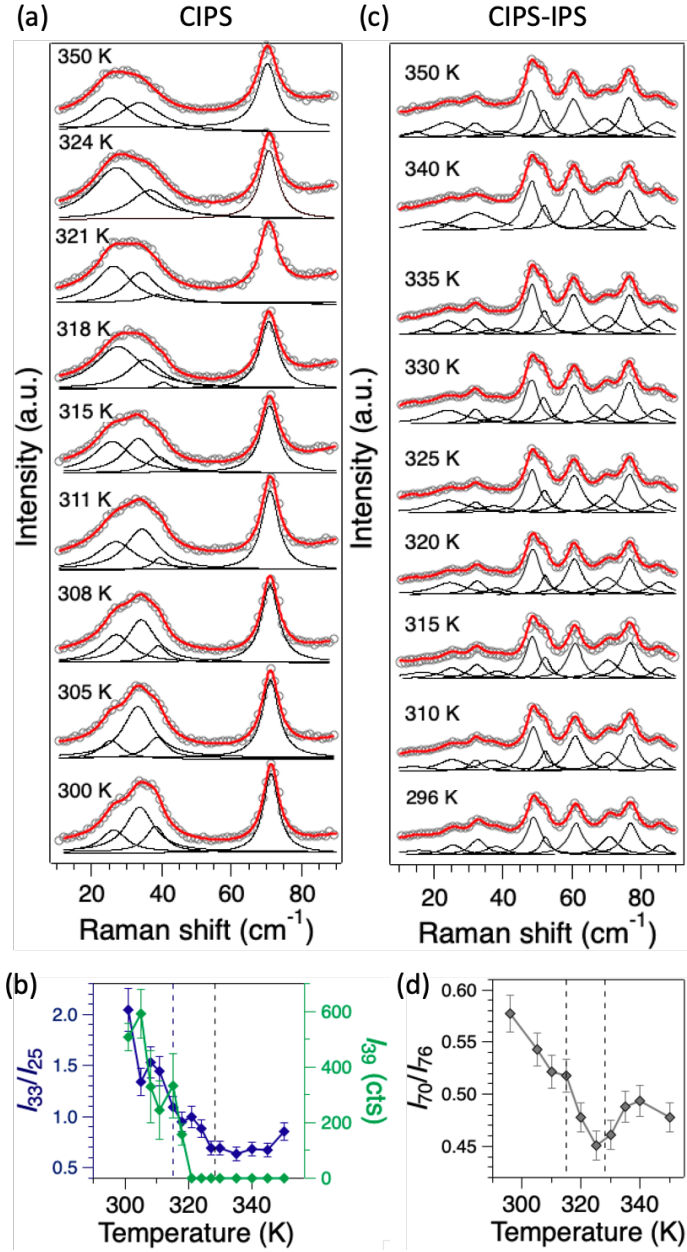

FIG. 2. Temperature-dependent Raman spectra across the ferrielectric-paraelectric phase transition in (a) CIPS and (c) CIPS-IPS crystals. (b) Intensity ratio of the copper vibrational modes around 33 and 25 cm$^{-1}$ ($I_{33}/I_{25}$, left axis) and the intensity of the 39 cm$^{-1}$ peak ($I_{39}$, right axis) against temperature in CIPS (d) Intensity ratio of the copper vibrational modes at 70 and 76 cm$^{-1}$ ($I_{70}/I_{76}$) against temperature in CIPS-IPS.

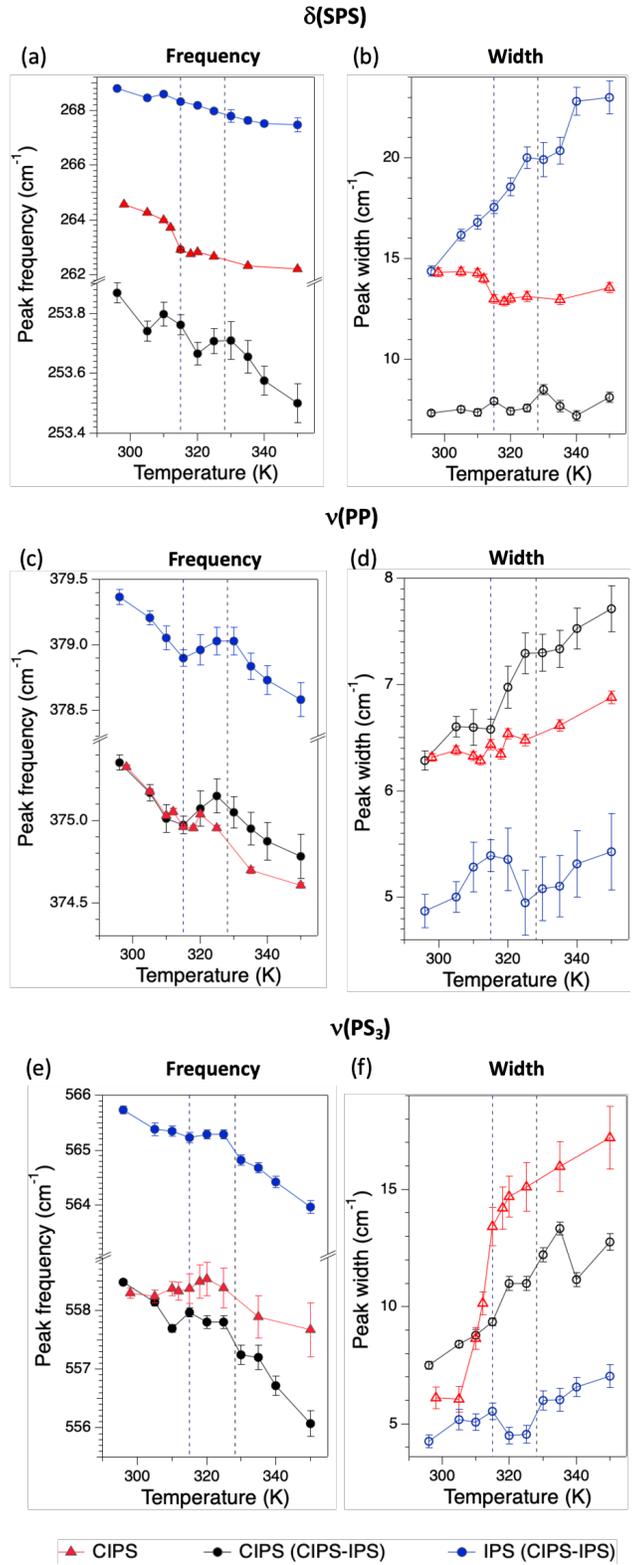

FIG 3. Temperature-dependent frequencies (filled data points) of the (a) $\delta$(SPS), (c) $\nu$(PP), (e) $\nu$(PS)$_3$ peaks and widths (hollow data points) of the (b) $\delta$(SPS), (d) $\nu$(PP), (f) $\nu$(PS$_3$) peaks.

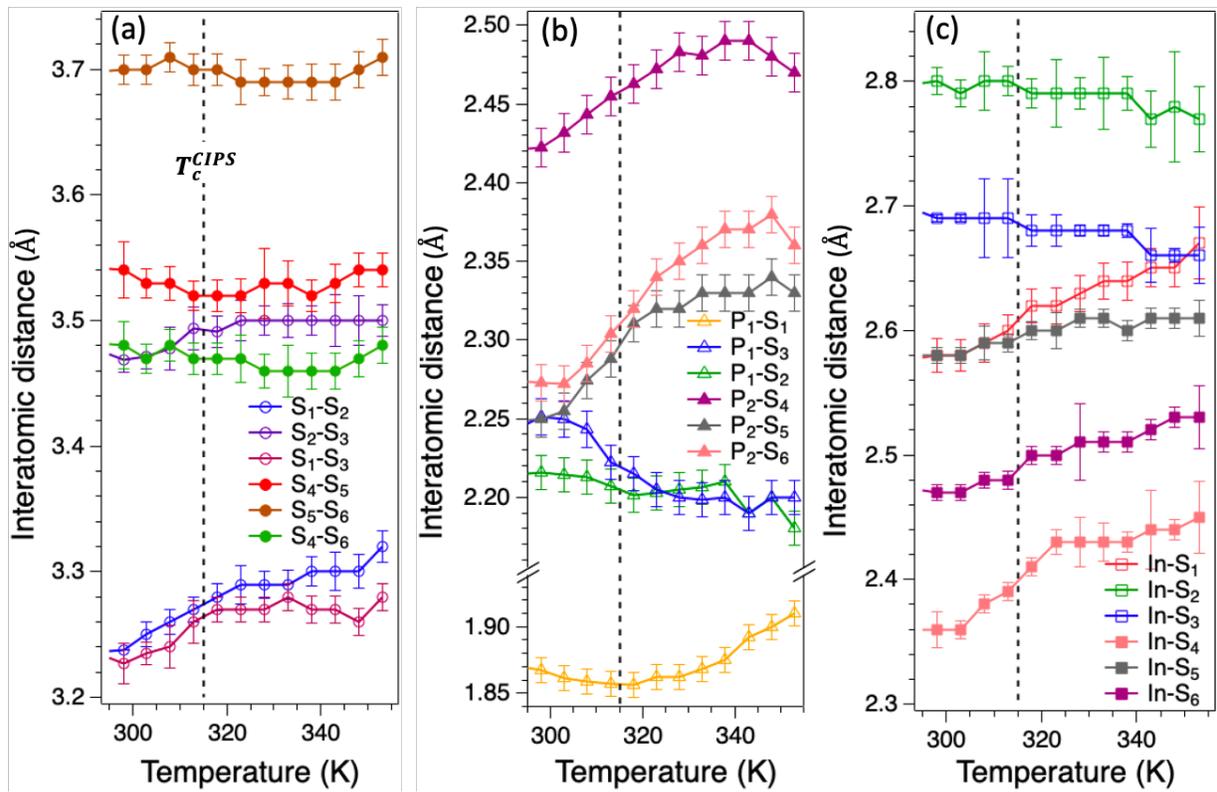

FIG. 4. Temperature-dependent distances between (a) S-S, (b) P-S, and (c) In-S atoms, calculated from Rietveld refinement of the XRD data. The data corresponding to the top and bottom of the $S_6$ octahedra are shown as open and filled data, respectively.